\documentclass[submission,Proceedings]{SciPost}

\hypersetup{
    colorlinks,
    linkcolor={red!50!black},
    citecolor={blue!50!black},
    urlcolor={blue!80!black}
}

\usepackage{csquotes}
\usepackage[bitstream-charter]{mathdesign}
\usepackage{siunitx}

\DeclareSymbolFont{usualmathcal}{OMS}{cmsy}{m}{n}
\DeclareSymbolFontAlphabet{\mathcal}{usualmathcal}

\begin{document}

\begin{center}{\Large \textbf{
PineAPPL: NLO EW corrections for PDF processes\\
}}\end{center}

\begin{center}
C.\ Schwan\textsuperscript{1$\star$}
\end{center}

\begin{center}
{\bf 1} TIF Lab and Dipartimento di Fisica, Universit\`a degli Studi di Milano and INFN, Milan, Italy\\
* christopher.schwan@mi.infn.it
\end{center}

\begin{center}
\today
\end{center}


\definecolor{palegray}{gray}{0.95}
\begin{center}
\colorbox{palegray}{
  \begin{tabular}{rr}
  \begin{minipage}{0.1\textwidth}
    \includegraphics[width=22mm]{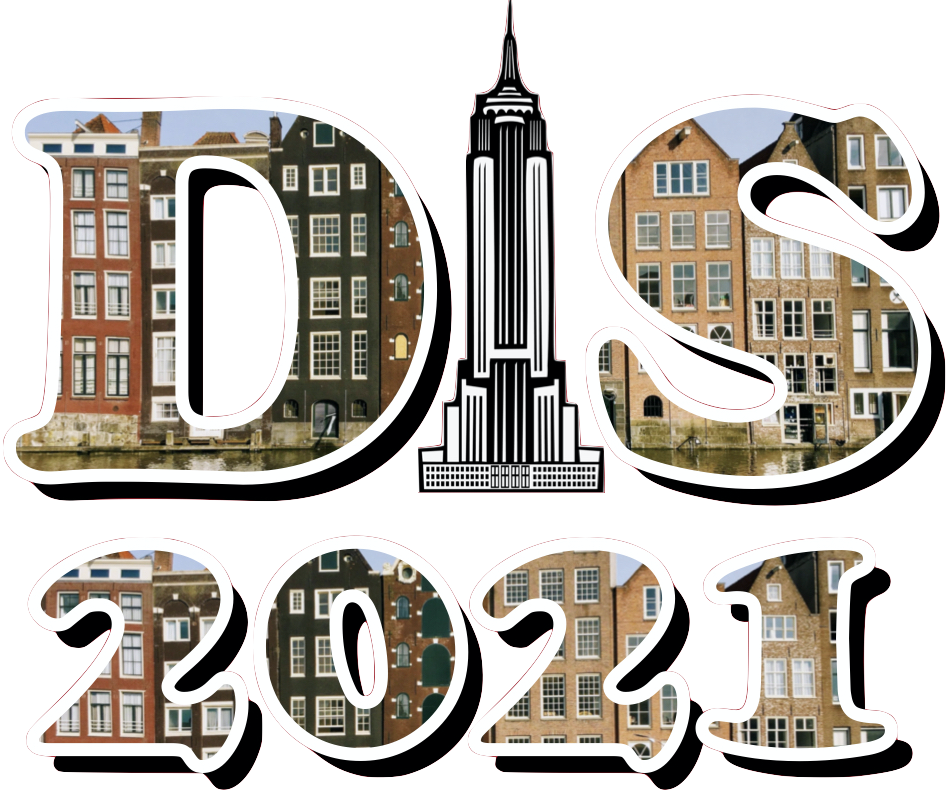}
  \end{minipage}
  &
  \begin{minipage}{0.75\textwidth}
    \begin{center}
    {\it Proceedings for the XXVIII International Workshop\\ on Deep-Inelastic Scattering and
Related Subjects,}\\
    {\it Stony Brook University, New York, USA, 12-16 April 2021} \\
    \doi{10.21468/SciPostPhysProc.?}\\
    \end{center}
  \end{minipage}
\end{tabular}
}
\end{center}

\section*{Abstract}
{\bf
Modern parton distribution function (PDF) determinations either neglect higher-order corrections in the electroweak (EW) coupling or implement them approximately for a subset of PDF processes.
We present a new tool, \textsc{PineAPPL}, which supports perturbative correction in arbitrary powers of the EW and strong coupling constant, and stores theoretical predictions independently of the used PDFs in so-called interpolation grids.
These are the foundation of any PDF determination, and will allow us to perform a PDF fit taking into account EW corrections for all processes consistently.
Apart from PDF determinations \textsc{PineAPPL} can also be used in precision phenomenology to study the impact of different choices of PDF sets, without rerunning the time-consuming computation.
As an application of this tool we show Drell--Yan (DY) lepton-pair production were larger effects of NLO EW corrections can be seen that will potentially influence the PDF fit and comment about which observables are particularly suited and NLO EW PDF fit and which are not.
}

\section{Introduction}

PDF determinations (for a recent review see e.g.\ Ref.~\cite{Ethier:2020way}) typically neglect the effects of higher-order effects in the EW coupling constant, $\alpha$, and only include corrections in the strong coupling constant $\alpha_\mathrm{s}$, up to next-to-next-to-leading order (NNLO) in recent PDF fits; a notable exception is the MSHT20~\cite{Bailey:2020ooq} PDF set, which however uses K-factors that approximately implement the effect of EW corrections for a subset of PDF processes.
With uncertainties of about \SI{5}{\percent} --- of the PDFs themselves, and by convolution also the cross-sections --- neglecting/approximating NLO EW effects is certainly justified, given that the size of EW corrections are often of the size of \SI{1}{\percent} or even smaller.
However, it is foreseeable that uncertainties of PDFs will be much smaller in the future, so that the task of including EW corrections in a PDF fit becomes increasingly important.
Apart from this more general statement we note the following specific observations: 1) in phase-space regions where many events are recorded from the experiments the PDFs are typically much more precise, so that even small EW correction can become important; 2) for more extreme phase-space regions, for example large transverse momentum of the $\mathrm{Z}$ boson, EW corrections can be of the order of \SI{-20}{\percent} data are still measured.
At the LHC for run I and II the datapoints typically have large experimental uncertainties to accommodate potentially large theoretical corrections; however, this is expected to change in run III.
Finally we note that including EW corrections in a PDF fit requires a more consistent treatment of data, which in practice means not subtracting collinear photons, which is typically done for a large number of gauge-boson production datasets at the LHC, for which this procedure can have a sizeable impact.

\section{PineAPPL: EW corrections and interpolation grids}

In this article we use the term \enquote{EW corrections} loosely and let it refer to any perturbative order that is not purely of a strong origin.
For PDF processes this includes, for example, the $\mathcal{O} (\alpha^3)$ for DY (with decays), which is a purely EW correction, but for jet production we also include the additional leading orders at $\mathcal{O} (\alpha_\mathrm{s} \alpha)$ and $\mathcal{O} (\alpha^2)$, and the NLOs at $\mathcal{O}(\alpha_\mathrm{s}^2 \alpha)$, $\mathcal{O}(\alpha_\mathrm{s} \alpha^2)$, which we sometimes call combined QCD--EW corrections, and finally the purely EW correction of $\mathcal{O} (\alpha^3)$.

Including these orders into a PDF fit requires the calculation of interpolation grids, which are essentially cross sections of processes differential with respect to the parton momentum fractions $x_1$ and $x_2$, their initial-state flavours $a$ and $b$ and finally the choice of the scale $Q^2$, in separate powers of $\alpha_\mathrm{s} (Q^2)$ and $\alpha$:
\begin{equation}
\frac{\mathrm{d} \sigma_{ab}}{\mathrm{d} \mathcal{O}} (x_1, x_2, Q^2) = \sum_{k,l,m,n} \alpha_\mathrm{s}^k \left( Q^2 \right) \alpha^l \log^m (\xi_\mathrm{R}^2) \log^n (\xi_\mathrm{F}^2) W_{ab}^{(k,l,m,n)} \left( x_1, x_2, Q^2, \mathcal{O} \right) \text{.}
\label{eq:interpolation-grids}
\end{equation}
The scale-variation grids, which have $m > 0$ or $n > 0$, are important when the user wants to vary the renormalisation, $\mu_\mathrm{R}$, and factorisations scale, $\mu_\mathrm{F}$, around the central scale choice $Q^2$ using the ratios $\xi_\mathrm{R}^2 = \mu_\mathrm{R}^2/Q^2$ and $\xi_\mathrm{F}^2 = \mu_\mathrm{F}^2/Q^2$.

These grids are the preferred representation of theory predictions for PDF fits, but outside of PDF fits they are also useful to quickly study the impact of different PDF sets to observables.
They can be generated once, and after their generation convolutions with arbitrary PDFs are performed in a matter of seconds.

An alternative to interpolation grids are K-factors, in our case EW K-factors,
\begin{equation}
K_\text{EW} = \frac{\mathrm{d} \sigma_\text{EW}/\mathrm{d} \mathcal{O}}{\mathrm{d} \sigma_\text{QCD}/\mathrm{d} \mathcal{O}} \text{,}
\end{equation}
which are the ratios of cross sections with EW (and NNLO QCD) corrections to the cross section without EW corrections, only including higher orders in QCD.
Eq.~\eqref{eq:interpolation-grids} is then approximated using
\begin{equation}
\frac{\mathrm{d} \sigma_{ab}}{\mathrm{d} \mathcal{O}} (x_1, x_2, Q^2) \approx K_\text{EW} \sum_k \alpha_\mathrm{s}^k \left( Q^2 \right) \alpha^{l_\mathrm{min}} W_{ab}^{(k,l_\mathrm{min})} \left( x_1, x_2, Q^2, \mathcal{O} \right) \text{,}
\end{equation}
where $l_\mathrm{min}$ is the smallest power of $\alpha$ for the process at LO.
The advantage of the method is that an NNLO QCD PDF fit can be easily upgraded to approximately include NLO EW effects (see for example Refs.~\cite{Boughezal:2017nla,AbdulKhalek:2020jut,Bailey:2020ooq}), however, there are several pitfalls stemming from the fact that the K-factor is applied after the PDF convolution (i.e.\ is a hadronic K-factor and not a partonic one): it is blind to the initial-state flavours, the partonic momentum fractions and the scale.
For example, it is possible that at NLO EW a new partonic channel opens up, typically $\mathrm{q}\gamma$, which is positive and cancels negative corrections of a virtual correction.
This results in small K-factors, which leave the PDF fit unaffected, and therefore underestimating the EW corrections.
To be on the safe side, especially for the accuracy mentioned above, we therefore take EW corrections into account using interpolation grids instead of K-factors.

To this end we developed \textsc{PineAPPL}~\cite{Carrazza:2020gss}, which is an interpolation library~\cite{christopher_schwan_2021_4636076}, similar to \textsc{APPLgrid}~\cite{Carli:2010rw} and \textsc{fastNLO}~\cite{Kluge:2006xs,Wobisch:2011ij,Britzger:2012bs}, but it supports arbitrary perturbative predictions in powers of $\alpha_\mathrm{s}$ and $\alpha$ as shown in Eq.~\eqref{eq:interpolation-grids}.
While extending the two previously mentioned interpolation libraries would have been possible in principle, we found that the increased memory requirements for supporting the additional orders in $\alpha$ made a rewrite from scratch inevitable in practice.

\textsc{PineAPPL} is the technical solution that will allow us to generate the interpolation grids needed for PDF fits including EW corrections.
It consists of a library, written in Rust~\cite{10.1145/2663171.2663188}, which we interfaced to \textsc{Madgraph5\_aMC@NLO}~\cite{Alwall:2014hca,Frederix:2018nkq}, similarly to \textsc{aMCfast}~\cite{Bertone:2014zva}, which it replaces for \textsc{Madgraph5\_aMC@NLO} version 3 and higher.
In addition to a library a command-line program \texttt{pineappl} is provided, which allows for quick and easy convolutions of the interpolation grids with arbitrary PDF sets through \textsc{LHAPDF}~\cite{Buckley:2014ana}, evaluations of the PDF uncertainties, scale variations and plotting.
An example output of the plotting command is shown in Fig.~\ref{fig:drell-yan-plots}.
See also Sec.~\ref{sec:conclusion} for instructions on how to produce a similar plot using the toolchain described above.

\section{CMS Drell--Yan lepton-pair production at 13~TeV}
\label{sec:cms-dy-production}

\begin{figure}
\centering
\includegraphics[width=0.95\textwidth]{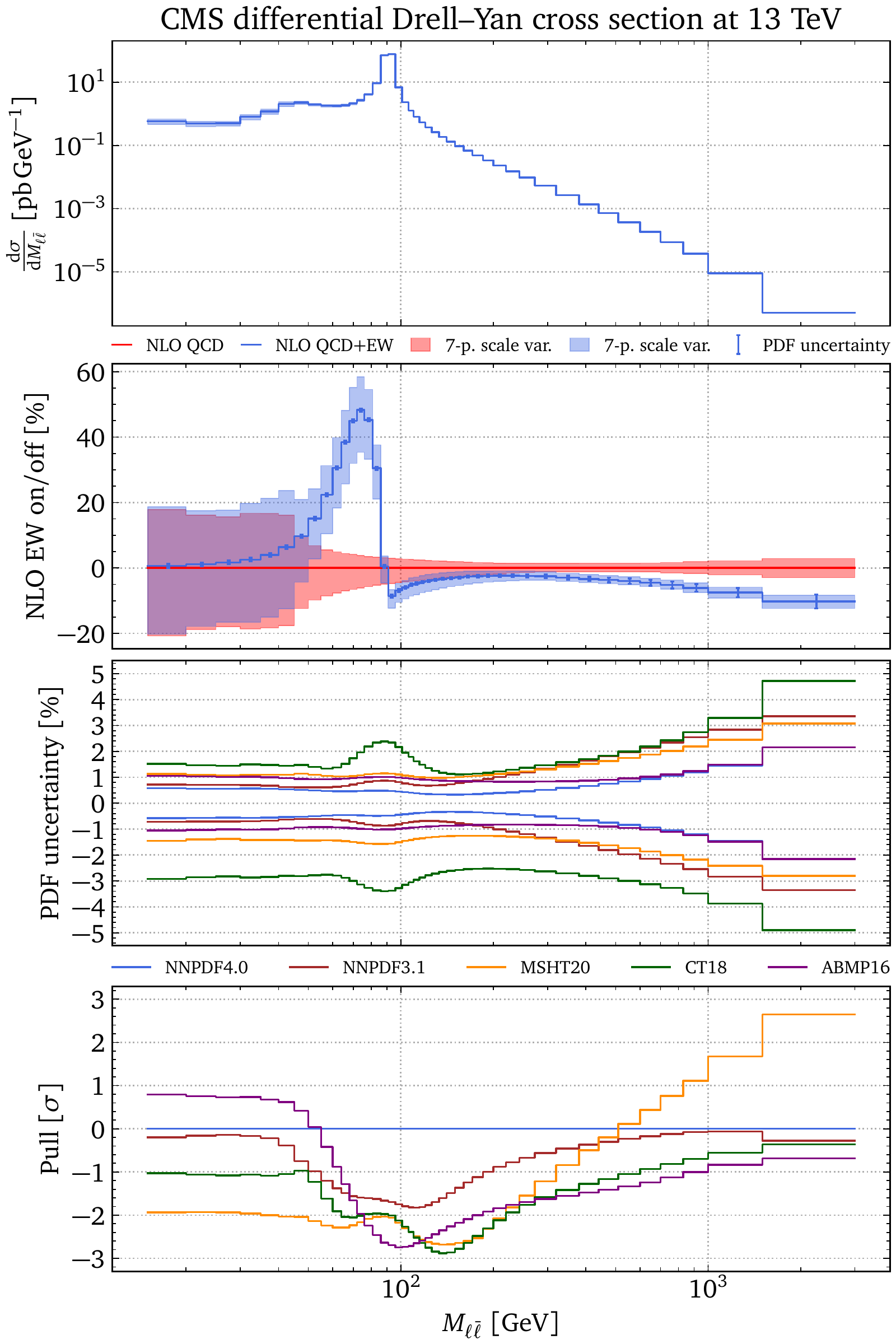}
\caption{Theoretical predictions for the CMS DY lepton-pair production at \SI{13}{\tera\electronvolt}.
Top panel: absolute prediction for the cross section differentially in the invariant mass of the lepton pair.
Second panel: NLO EW+QCD predictions relative to NLO QCD only.
Third panel: PDF uncertainties for the four global PDF sets NNPDF4.0 (preliminary fit), MSHT20, CT18 and ABMP16.
Bottom panel: Pull between NNPDF4.0 and the remaining three PDF sets.}
\label{fig:drell-yan-plots}
\end{figure}

As an example of the application of \textsc{PineAPPL} together with \textsc{Madgraph5\_aMC@NLO} v3.1 we show the theoretical prediction for the CMS measurement~\cite{CMS:2018mdl} of DY lepton-pair production at \SI{13}{\tera\electronvolt} in Fig.~\ref{fig:drell-yan-plots}.
This measurement shows the invariant mass $M_{\ell\bar{\ell}}$ of the lepton-pair very finely resolved around the $\mathrm{Z}$-boson peak, with bins as small as \SI{4}{\giga\electronvolt}.
As a result of this, QED effects produce very large higher-order corrections, which are part of the NLO EW corrections and can be seen in the second panel of Fig.~\ref{fig:drell-yan-plots}.
This effect is typically not so pronounced in the ATLAS and CMS measurements at \SIlist{7;8}{\tera\electronvolt}, where the bins are larger and are chosen to be symmetric around the peak, so that the corrections mostly cancel themselves out.

The second panel also shows the uncertainty band for the NLO QCD+EW correction (blue) and for the NLO QCD correction (red) only.
Note that the large corrections below the $\mathrm{Z}$-boson peak also enhance the uncertainties significantly (the largest difference is found for the range from \SIrange{76}{81}{\giga\electronvolt} where the uncertainty in the positive direction grows from \SI{3.5}{\percent} to \SI{6.4}{\percent}, which is an increase of \SI{83}{\percent}).
As a consequence including this dataset into a PDF fit with missing-higher order uncertainties (see e.g.\ Ref.~\cite{NNPDF:2019vjt,NNPDF:2019ubu}, but without this dataset) without NLO EW corrections and Born-level observables would seriously underestimate the theory uncertainties.

The third and fourth panel show the relative PDF uncertainties for NNPDF4.0 (preliminary fit, see Ref.~\cite{Rojo:2021gdq}) against NNPDF3.1~\cite{NNPDF:2017mvq}, MSHT20~\cite{Bailey:2020ooq}, CT18~\cite{Hou:2019efy} and ABMP16~\cite{Alekhin:2017kpj}, for which we chose the PDF sets with NNLO QCD accuracy and $\alpha_\mathrm{s} (M_\mathrm{Z}^2) = 0.118$.
The pull is calculated as
\begin{equation}
\operatorname{Pull} = \frac{X_2 - X_1}{\sqrt{(\delta X_1)^2 + (\delta X_2)^2}}
\end{equation}
where $X_1 = \mathrm{d} \sigma/\mathrm{d} M_{\ell\bar{\ell}}$ is the differential cross section calculated with the NNPDF4.0 PDF set and $X_2$ the same for any of the other sets, and $\delta X_1$ and $\delta X_2$ are the corresponding PDF uncertainties.
The pull of each PDF with respect to NNPDF4.0 is below $2 \sigma$ for most of the PDFs and for most of the invariant mass, however, around the $\mathrm{Z}$-boson peak the pull is as large as $3 \sigma$, also due to the very small uncertainties of the NNPDF4.0 PDF set.
In this region we expect the inclusion of theoretical uncertainties to produce larger PDF uncertainties in NNPDF4.0, and in turn smaller pulls.
However, given that the pull between NNPDF4.0 and 3.1 are always below $2 \sigma$ shows that the two PDF sets are consistent with each other.

\section{Lepton observables for different theoretical predictions}

\begin{figure}
\centering
\includegraphics[width=0.4\textwidth]{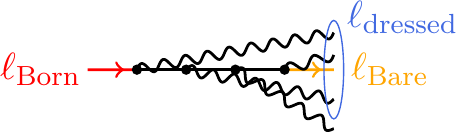}
\caption{Different lepton observables for which data are published: 1) Born-level leptons, and 2) bare leptons, which are leptons before and after radiation, respectively.
Finally, 3) dressed-leptons, whose momenta include the surrounding photons.}
\label{fig:collinear-photon-off-a-lepton}
\end{figure}

As can be seen in the second panel of Fig.~\ref{fig:drell-yan-plots} the NLO EW corrections are very large and distort the shape of the $M_{\ell\bar{\ell}}$ distribution around the Z-boson peak.
This effect is of course well known to come from soft photons, and therefore measurements of leptonic observables are typically published at \emph{Born-level}, as shown in Fig.~\ref{fig:collinear-photon-off-a-lepton}.
Born-level leptons are a measurement combined with a theoretical prediction that removes the effect from soft photons (typically obtained with PHOTOS~\cite{Golonka:2005pn,Barberio:1993qi,Barberio:1990ms}).
The advantage of using leptons \enquote{before they radiate} is that the large corrections from soft photons are taken care of and that these observables can be compared to pure QCD predictions with good accuracy, which is what is done in most global PDF fits.
However, if we want to take into account weak corrections, which are important e.g.\ in DY in the tail of the $M_{\ell\bar{\ell}}$ distribution, Born-level leptons pose the problem that some QED effects have been subtracted from the data, but the rest of the EW corrections have not been.
A comparison of data against theory with full EW corrections is therefore not possible without having some sort of double-counting, and therefore we have to use different lepton-observables, the most convenient which are \emph{dressed-lepton observables} (see Fig.~\ref{fig:collinear-photon-off-a-lepton}).
Dressed-lepton observable combine the momenta of leptons with the momenta of photon around a certain radius, where typically $\Delta R = 0.1$ is chosen.
For the observables both measurements and predictions can be easily done.

\section{Conclusion}
\label{sec:conclusion}

We discussed the aim of performing a PDF fit with the consistent inclusion of EW corrections, for which we wrote the interpolation library \textsc{PineAPPL}.
Together with \textsc{Madgraph5\_aMC@NLO}, to which it is interfaced, we are able to produce predictions for all PDF processes including the EW corrections, and we discussed an application for the CMS DY lepton-pair production, showcasing features of the EW corrections: 1) large and shape-changing QED corrections, 2) enlarged theory uncertainties and 3) large and negative weak corrections in the tails of the invariant mass distribution.
These predictions, when compared to measurements require an appropriate definition of lepton observables --- dressed leptons --- which we discussed is different from the ones used for purely NNLO QCD predictions.

Finally, we note that we published a script~\cite{bash_script} can be used to produce a plot similar to the one shown in Fig.~\ref{fig:drell-yan-plots} and which gives an in-depth showcase of how the toolchain is used and what it can accomplish.
Before running the script \textsc{Madgraph5\_aMC@NLO} v3.1 or higher (see \url{https://launchpad.net/mg5amcnlo}) and of PineAPPL v0.4.1 or higher (see \url{https://github.com/N3PDF/pineappl}) must be installed.

\paragraph{Funding information}
C.S.\ is supported by the European Research Council under the European Union's Horizon 2020 research and innovation Programme (grant agreement no.\ 740006).

\bibliography{article.bib}

\nolinenumbers

\end{document}